\documentclass[%
 reprint,
 amsmath,amssymb,
 aps,
pra,
floatfix,
]{revtex4-2}

\usepackage{graphicx}
\usepackage{dcolumn}
\usepackage{bm}
\usepackage{siunitx}
\usepackage{braket}

\DeclareSIUnit\torr{Torr}

\newcommand{\crossoverBraKet}{$\ket{{^1S_0}, F = 1/2} \leftrightarrow \ket{{^1P_1}, F' = 1/2, 3/2}$}

\newcommand{\motTransitionBraKet}{$\ket{{^1S_0}, F = 1/2} \leftrightarrow \ket{{^1P_1}, F' = 3/2}$}

\newcommand{\yibbyOneSevenOne}{$^{171}$Yb}

\newcommand{\yibbyOneSevenFour}{$^{174}$Yb}

\newcommand{\blueTransition}{${^1S_0} \leftrightarrow  {^1P_1}$}

\usepackage[colorlinks=true, allcolors=blue]{hyperref}

\begin{document}

\title{Demonstration of a simple and compact ytterbium magneto-optical trap}

\author{Benjamin White$^{1}$}
\author{Rachel F. Offer$^{1}$}
\author{Ashby P. Hilton$^{1}$}
\author{Elizaveta Klantsataya$^{1}$}
\author{Christopher J. Billington$^{1}$}
\author{Nicolas Bourbeau H\'{e}bert$^{1}$}
\author{Montanna Nelligan$^{1}$}
\author{Andre N. Luiten$^{1}$}

\affiliation{$^{1}$Institute for Photonics and Advanced Sensing (IPAS) and School of Physics, Chemistry and Earth Sciences, University of Adelaide, Adelaide, SA 5005, Australia.}

\date{\today}

\begin{abstract}
We present a low Size, Weight and Power (SWaP), low-complexity, ytterbium magneto-optical trap (MOT). We demonstrate trapping of \num{1.4e6}\,\yibbyOneSevenOne\,atoms on the \motTransitionBraKet\, transition directly from a hot thermal beam. We explore the effect of trap detuning and oven temperature on trap number, density, loading rate and sample temperature. The low SWaP and low-complexity design presents a realistic pathway towards portable ytterbium MOTs, allowing cold atom ytterbium systems to escape the confines of the laboratory and perform precision measurements in field environments. 
 
\end{abstract}

\maketitle



\section{\label{sec:level1}Introduction}

The introduction of laser cooling in the 1970s revolutionized atomic physics by enabling scores of precision measurement technologies and fundamental studies \cite{Raab1987, Chu1998, Tannoudji1998, Phillips1998, Grimm2000}. Among these, the magneto-optical trap (MOT) remains an essential technology for confining and probing neutral atoms. The flexibility of the atomic structure of group II-like elements, such as ytterbium (Yb), is especially attractive for next-generation cold-atom technologies. Ground-state transitions in Yb support both primary (\SI{399}{\nano\meter}) and secondary (\SI{556}{\nano\meter}) cooling as well as an ultra-narrow line (\SI{578}{\nano\meter}) for precision measurements, making it a strong candidate for optical atomic clocks \cite{Ludlow2015, Schioppo2017, Gao2018,  White2024, Hilton2025}, quantum sensing \cite{Covey2023,Shin2024}, quantum simulation \cite{Gross2017, Taie2022, Jia2024} and fundamental physics research \cite{Hudson2011,Dzuba2018,Tang2023,Door2025}. 

Presently Yb-based cold atom systems are mostly confined to laboratory settings \cite{Schioppo2017, Gao2018, Letellier2023, Nomura2023} due to the size and complexity of the laser cooling apparatus. The high temperature required to produce a modest vapor number density (typically $\geq$ \SI{400}{\degreeCelsius}) along with the need for short-wavelength laser sources often results in large, complex and power intensive setups. Moreover, the broad \SI{399}{\nano\meter} transition necessitates large magnetic field gradients, hence large currents, which often involve active cooling of the coils. Further, most Yb MOTs rely on Zeeman slowing \cite{Honda1999a, Honda2002, Letellier2023, Pandey2010a, Wodey2021, Loftus2000, Belotelov2020} or far-detuned slowing beams \cite{Park2003, Takasu2003,Kawasaki2015, Nomura2023} to maximize atom number, adding further size and complexity. There has been recent interest in the fabrication of optical elements that create the electromagnetic fields and polarizations required for a MOT from a single input beam using both diffraction gratings (gMOT) \cite{Fasano2021, Sun2021,  Vyalykh2024}, and reflectors \cite{Bondza2024,Pick2024}. However, the challenge of building a compact physics package still remains. 

Here, inspired by early work showing that robust Yb MOTs can be realized without dedicated longitudinal cooling apparatus \cite{Loftus2000,Rapol2004,Loftus2001a}, we report on a compact, low-complexity, low Size, Weight and Power (SWaP) Yb MOT. These foundational experiments demonstrate that by placing a thermal Yb source in close proximity to the trapping region, a sufficient atomic flux can be captured directly from the background vapor negating the need for a Zeeman slower \cite{Loftus2000,Rapol2004}. This simple configuration was also shown to be effective for the simultaneous trapping of multiple Yb isotopes, highlighting the versatility of the approach \cite{Loftus2001a}. We build on this work and further simplify the system by using a single commercial laser source for both trapping and diagnostics, a low-complexity atomic source, and trapping atoms within an off-the-shelf, uncoated glass vacuum cell. Operating on the \motTransitionBraKet \, transition, our apparatus traps \num{1.4e6}{\,atoms}, at a density of \num{3.1e9}{\unit{\,atoms\per\centi\meter\cubed}} and achieves a minimum temperature of \num{680(90)}{\,\unit{\micro\kelvin}. We investigate the dependence of MOT characteristics: temperature, atom number, trap density, and loading rate, on cooling laser detuning and oven temperature. This low SWaP, low-complexity MOT design demonstrates a pathway for the deployment of the next generation of portable Yb cold atom experiments. 

\begin{figure*}
{\includegraphics{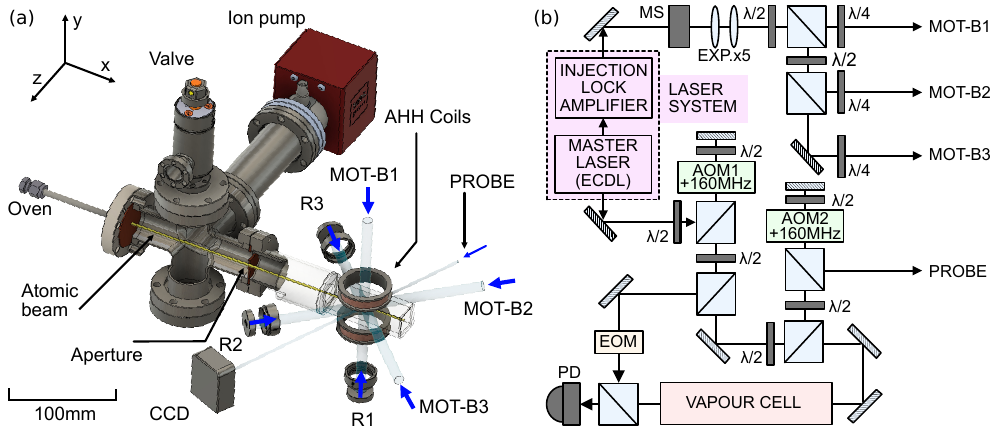}}
\caption{(a) 3D model of the physics package. Overall length is \SI{440}{\milli\meter}. Here, MOT-B1, MOT-B2 and MOT-B3 refer to incoming MOT cooling beams. R1, R2 and R3 refer to each retro-reflected beam's mirror and quarter-wave plate. AHH refer to the anti-Helmholtz coil pair. Position and orientation shown for the probe light and CCD camera used for all diagnostics. Cut away shows the path of the thermal atomic beam within the vacuum system. Axes are defined for reference and consistency with results. (b) Optical layout. Single laser system (commercial diode and injection lock amplifier) responsible for three different frequencies of light for the MOT: Locking, trapping and probing. MS: Mechanical shutter, EXP: beam expansion lenses, AOM: acousto-optic modulator, EOM: electro-optic modulator, PD: photo-detector,  $\lambda$/2: half wave plate,  $\lambda$/4: quarter-wave plate. MOT-B1, MOT-B2, MOT-B3 refer to the three input cooling beams for the MOT, PROBE is resonant light used for absorption measurements.}
\label{physics_and_optics} 
\end{figure*}

\section{Physics Package}

The physics package, shown in Fig.\,\ref{physics_and_optics}(a), contains the vacuum system, including the atomic source, ion pump and spectroscopy cell, as well as the coils and imaging setup. It is entirely constructed from off-the-shelf vacuum components, measuring \SI{440}{\milli\meter} in length, \SI{300}{\milli\meter} in depth and \SI{250}{\milli\meter} in height, weighing \SI{15}{\kilo\gram} and requiring approximately \SI{50}{\watt} of power. The atomic source is a Knudsen tube design comprised of several Swagelok brand pipe fittings from the SS316 dual ferrule range, mounted to a standard CF 2.75” flange adapter with a stainless-steel tube \SI{150}{\milli\meter} long, \SI{3}{\milli\meter} inner diameter. Although not rated for Ultra-high vacuum (UHV), we find that after careful cleaning and bake-out, these fittings provide an excellent seal that allows us to maintain \SI{e-9}{\torr} operation over many thermal cycles. The reservoir contains a \SI{1}{\gram} pellet of high purity Yb, which has lasted \num{3} years of daily use. The outer surface of the oven is wound with fiberglass insulated Nichrome heater wire, and heated to an operating temperature between \SI{350}{\degreeCelsius} to \SI{450}{\degreeCelsius}. Oven temperature is monitored by a thermocouple mounted between the stainless tube and the heater wire. The collimation of the atomic beam  of \SI{20}{\milli\radian} is set by the aspect ratio of the oven geometry, then further constrained to \SI{10}{\milli\radian} by a \SI{3}{\milli\meter} diameter aperture in a solid copper gasket mounted between the 5-way cross and the glass spectroscopy cell. This copper gasket also has an additional slot that allows for improved vacuum conductance while protecting the interior cooling windows from line of sight of the oven. The trapping region is within a rectangular glass cell which allows \SI{270}{\degree} optical access to the MOT. A combined ion pump and non-evaporable getter maintains a pressure of \SI{1e-9}{\torr} within the vacuum system and provides a proxy measurement of the background pressure.

The quadrupole magnetic field required for the MOT is produced by a pair of anti-Helmholtz configuration magnetic coils. Each coil consists of \num{20} turns of \SI{2.5}{\milli\meter} copper wire with an inner diameter of \SI{30}{\milli\meter} and separation of \SI{30}{\milli\meter}. A DC power supply drives both coils in series with \SI{20}{\ampere} of current, producing a field gradient of \SI{52}{G\per\centi\meter} with a power dissipation of \SI{10}{\watt} per coil. The coils are soaked in a lacquer that helps hold their shape without a separate former, and also aids in thermal conduction out of the center of the coils, and thus no active cooling of the coils is required during extended (\num{>3}\,hours) operation of the MOT. The glass cell allows us to position our coils in close proximity to the trapping region, making large magnetic field gradients easier to produce with less current draw. The use of a glass cell also negates the effect of eddy currents when switching magnetic fields (not implemented here) \cite{Letellier2023}.

\section{Optics}
Figure\,\ref{physics_and_optics}(b) shows a schematic of the optical components used to generate all of the MOT light. Operation of the MOT requires three frequencies of \SI{399}{\nano\meter}: $\mathrm{\omega_{MTS}}$ for modulation transfer spectroscopy for frequency stabilization, $\mathrm{\omega_{MOT}}$ for the MOT cooling beams, and $\mathrm{\omega_{PROBE}}$ used for the MOT measurements. All light originates from an external-cavity diode master laser (ECDL). Figure\,\ref{energy_levels} shows these three frequencies of light generated in reference to the \yibbyOneSevenOne\, \crossoverBraKet\, transitions. 

MOT cooling beams are produced by an injection-locked amplifier (ILA) seeded by the ECDL. The ILA is capable of generating $>$\SI{200}{\milli\watt} of \SI{399}{\nano\meter} light. The output beam from the ILA passes through a mechanical shutter, used to extinguish trapping light. The beam's $1/e^2$ diameter is then expanded from \SI{2}{\milli\meter} to \SI{10}{\milli\meter}. The expanded beam is passed through two \SI{25}{\milli\meter} polarizing beam splitter (PBS) cubes to create three input MOT beams. These beams are retro reflected to create the six orthogonal trapping beams (Fig.\,\ref{physics_and_optics}(a)). The total power available for trapping is \SI{60}{\milli\watt} across the three MOT beams; corresponding to a per-beam saturation parameter $s_\mathrm{0} = 0.4$. A large fraction of the total laser power is lost due to the inefficient optical transmission of off-the-shelf optics at \SI{399}{\nano\meter}. The required circular polarization is set by quarter-wave plates before entry of the light to the glass cell and by quarter-wave plates placed in front of the retro-reflecting mirrors.

\begin{figure}[htbp]
\includegraphics{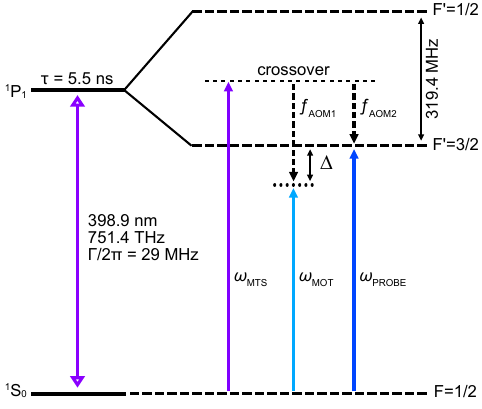}
\caption{Energy level diagram for the \yibbyOneSevenOne\, \motTransitionBraKet\,transition. Showing the frequency of the MTS lock ($\mathrm{\omega_{MTS}}$), the MOT cooling beams ($\mathrm{\omega_{MOT}}$), the laser detuning from resonance ($\Delta$), and the probe beam ($\mathrm{\omega_{PROBE}}$). Also shown are the excited state lifetime ($\tau$), natural linewidth ($\Gamma$), as well as the transition wavelength and frequency.}
\label{energy_levels}
\end{figure}

Light from the ECDL is also used for both laser stabilization and production of the diagnostic probe frequency. Initially light is shifted up via the double-passed AOM1 (Fig.\,\ref{physics_and_optics}(b)) by $2 \times $\SI{80}{\mega\hertz} plus $\Delta$, where $\Delta$ is the frequency detuning from resonance used for the trapping light ($\Delta \equiv \omega_{\mathrm{MOT}}-\omega_{\mathrm{res}}$). This light is then passed through a ytterbium vapor cell held at \SI{350}{\degreeCelsius}. Using modulation transfer spectroscopy (MTS) \cite{Preuschoff2018, Hilton2025}, we lock the AOM shifted light to the \crossoverBraKet \, crossover feature. The separation between the crossover and the \motTransitionBraKet \, transition is \SI{-159.7}{\mega\hertz}. Thus, the  AOM1 shift cancels the offset of the crossover feature, and the ILA output frequency is on resonance with the \motTransitionBraKet \, transition when $\Delta = 0$. By adjusting $\Delta$ we can choose a cooling offset between \SI{-60}{\mega\hertz} and \SI{+60}{\mega\hertz}.

Probe light for the MOT measurements is tapped off from the vapor cell light after AOM1, directly before the MTS spectroscopy measurement (Fig.\,\ref{physics_and_optics}(b)). The frequency of this light is on resonance with the crossover feature. This light is down shifted by \SI{160}{\mega\hertz} using the double-passed AOM2, placing it on resonance with the \motTransitionBraKet \,transition, with tunability of \SI{30}{\mega\hertz} either side of resonance, and allowing us to pulse the probe light if needed. The probe light is fiber coupled to the physics package with nominal output power of \SI{20}{\micro\watt}. Absorption imaging measurements are performed by passing probe light through the glass cell and through the MOT region onto a charge coupled device (CCD) camera, allowing characterization of the atomic sample. 

The setup is designed to target the \yibbyOneSevenOne\, isotope and is not tunable to other isotopes. We can however manually tune the laser frequency for the more abundant \yibbyOneSevenFour\, isotope and measure the loading time and atom number while the laser is unlocked and temporarily at an acceptable detuning, which we also report on.

\section{Results}

We characterize the atomic sample temperature, number and density and MOT loading rates. Unless otherwise noted, the default parameters used for these results are a magnetic field gradient of \SI{52}{G\per\centi\meter} and oven temperature of \SI{425}{\degreeCelsius}. Atom number and cloud temperature are measured using absorption (or shadow) imaging with the probe beam. To perform a measurement the MOT is loaded and a signal image (with atoms) taken with resonant probe light \SI{200}{\micro\second} after extinguishing the trapping light. A reference image (with no atoms) is taken \SI{100}{\milli\second} later once the atoms have fully dispersed. Analysis of these images is performed by calculating a transmission image, created by dividing the signal image by the reference image. The transmission is then converted to an optical depth ($OD$) map, which is fitted to a two-dimensional Gaussian function. The total atom number, $N$, is calculated by integrating the atomic column density, $n_{\mathrm{2D}}(x,y) = OD(x,y)/\sigma_{\mathrm{abs}}$, which for a Gaussian profile is given by:

\begin{equation}
N = \frac{OD_{\mathrm{peak}}}{\sigma_{\mathrm{abs}}}\;2\pi\,\sigma_{\mathrm{x}}\,\sigma_{\mathrm{y}}\,\sqrt{1-\rho^2}
\label{eq:AtomNumber}
\end{equation}

Here, $OD_{\mathrm{peak}}$ is the peak optical depth, $\sigma_{\mathrm{x}}$ and $\sigma_{\mathrm{y}}$ are the Gaussian widths, and $\rho$ is the correlation coefficient which accounts for any rotation of the cloud's principal axes relative to the imaging axes, all of which are extracted from the fit. The constant $\sigma_{abs}$ is the on-resonance absorption cross-section for the \motTransitionBraKet\, transition, taken as \SI{6.8e-14}{\meter\squared}\cite{Steck2024,foot2005atomic}. The temperature is determined from the evolution of the cloud size over the expansion time. We collect ten runs per data set and use the mean and standard error of the mean as the value and error bars in the following figures. 

The relatively high MOT temperatures measured make it difficult to switch the magnetic fields off before the cloud has dispersed. Rather than using high speed switching circuits, we instead leave the magnetic field gradient on during imaging. The Land\'{e} g-factor for the cycling transition is $g_{\mathrm{F}} = 2/3$, which corresponds to a frequency shift of \SI{1.4}{\mega\hertz\per G}. As such, the region within \SI{2}{\milli\meter} of the zero of the field gradient experiences a Zeeman shift of less than \SI{14.5}{\mega\hertz}, or $\Gamma/2$, where $\Gamma$ is the natural linewidth of the \blueTransition\,transition. By restricting our cloud expansion to diameters smaller than this, we can ensure that the Zeeman effect does not strongly alter our measurements of MOT size or atom number.

\begin{figure}[b]
\includegraphics{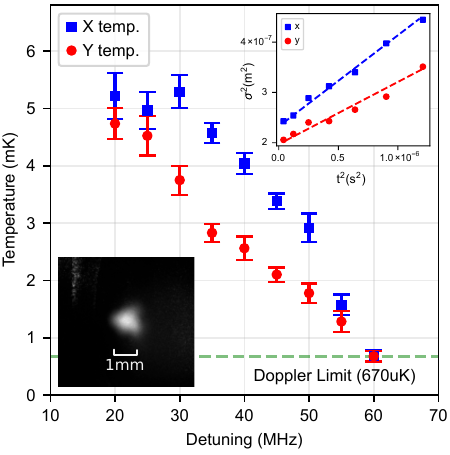}
\caption{Main: Temperature of the atomic sample as a function of trapping laser detuning in $x$ and $y$ directions. Error bars represent the standard error of the mean over \num{10} runs. Inset top: Example of temperature fit to MOT expansion over time. Where $\sigma_{x/y}$ represents the Gaussian width of the atomic cloud in the $x$ and $y$ direction, and the dashed line a linear fit to the cloud expansion. Inset bottom: CCD image of the MOT.}
\label{temperaturePlot}
\end{figure}

Figure\,\ref{temperaturePlot} presents the dependence of MOT temperature on trapping laser detuning. Both the $x$ and $y$ directions exhibit a similar trend: as the detuning increases from \qtyrange[]{20}{60}{\mega\hertz}, the temperature decreases monotonically. This behavior is consistent with Doppler cooling theory, which predicts a lower final temperature for larger detunings due to the reduced scattering rate \cite{Metcalf2003}. This has been seen previously in ytterbium \cite{Kostylev2014a}, and strontium \cite{Xu2003a}. While the temperature in both directions follow the same general trend, a noticeable temperature anisotropy exists between the $x$ and $y$ directions, particularly between \qtyrange[]{30}{50}{\mega\hertz}. This anisotropy is primarily set by the MOT quadrupole. In our geometry the field gradient along $y$ is about twice that along $x$ $\bigl(\partial B/\partial y \approx 2\,\partial B/\partial x\bigr)$, which yields different restoring forces and damping and thus different steady-state temperatures. With the coils left on during time-of-flight, a residual magnetic force $F=\mu\,\partial B/\partial y$ can bias the ballistic expansion along the strong-gradient axis. Secondary contributions include imbalances in laser intensity or polarization between the counter-propagating beams along the different axes or imperfections in optical beams resulting in spatial intensity variations, and leading to additional heating processes \cite{Chaneliere2005, Vogel1999,Xu2002,Xu2003a}. At maximum detuning (\SI{60}{\mega\hertz}), sub-millikelvin temperatures are achieved, approaching the Doppler limit of \SI{673}{\micro\kelvin} for the \blueTransition\, transition.

\begin{figure}[t]
\includegraphics{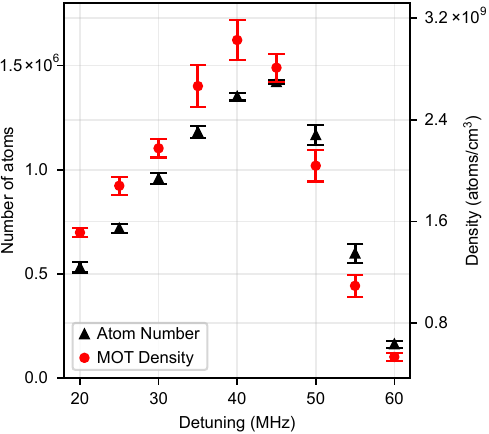}
\caption{Left y-axis (Black triangles): Number of atoms trapped as a function of trapping laser detuning. Right y-axis (Red Circles): Atomic density of trapped atoms as a function of trapping laser detuning. Error bars represent the standard error of the mean over \num{10} runs.}
\label{numberDensity}
\end{figure}

Figure\,\ref{numberDensity} presents the atom number and density as a function of cooling light detuning. The MOT atom number and density peak at \SI{45}{\mega\hertz} and \SI{40}{\mega\hertz} respectively, and then rapidly decrease with increasing detuning. This behavior can be understood from the interplay between several MOT mechanisms, which are outlined below.

For larger detunings ($|\Delta| > \Gamma$), the MOT's capture velocity, $v_{\mathrm{c}}$, scales roughly as $v_{\mathrm{c}} \propto |\Delta|/k$ (where $k$ is the cooling laser's wave-vector).  This allows the MOT to cool a larger fraction of velocity classes from the thermal atomic beam. However, the maximum scattering force, $F_\mathrm{scatt}$, simultaneously diminishes, scaling as $F_\mathrm{scatt} \propto s_{\mathrm{0}} \Gamma^2/\Delta^2$ for a constant saturation parameter $s_{\mathrm{0}}$. Conversely, at small detuning ($|\Delta|<\Gamma$), while the scattering force can be strong, the $v_{\mathrm{c}}$ range is significantly restricted. This limits the fraction of atoms from the thermal distribution that can be slowed and loaded into the MOT. Furthermore, operating very close to resonance will increase heating effects, limiting MOT stability and resulting in higher MOT temperatures (Fig.\,\ref{temperaturePlot}). For $|\Delta|=\SI{40}{\mega\hertz}$ we estimate $v_{\mathrm c}\approx\SI{24}{\meter\per\second}$~\cite{Metcalf2003}, far below the mean thermal speed ($\approx\SI{344}{\meter\per\second}$ at \SI{430}{\celsius}), so only a small fraction of atoms are capturable.

The observed optimum detuning represents a balance where the capture velocity is large enough to efficiently load atoms, while the scattering force remains sufficient for effective slowing, cooling, and confinement. The slight difference in the optimal detuning for atom number ($N$) versus density can be attributed to the detuning dependence of the MOT volume. The volume is influenced by the cloud temperature $T$, which, as seen in Fig.\,\ref{temperaturePlot}, decreases with increasing $|\Delta|$. Peak density may thus occur at a detuning that yields a particularly small MOT size due to low temperature, even if the atom number is not at its absolute maximum. In our case the most efficient cooling and trapping occurs at $\Delta\approx 3\Gamma/2$, this behavior is consistent with previous results for Yb MOTs loaded directly from thermal sources \cite{Rapol2004}.

Whilst the AOM shifts in our optical setup prevent stabilizing the laser at a frequency suitable for the \yibbyOneSevenFour\, isotope, we are able to take measurements by manually tuning the free running laser frequency to where we observe maximum fluorescence. We achieve an atom number of \num{4.1e6} \yibbyOneSevenFour\, (compared to \num{1.4e6} for \yibbyOneSevenOne), which is in agreement with previous work for \yibbyOneSevenOne\, \cite{Rapol2004, Park2003, Xu2009} and \yibbyOneSevenFour\,\cite{Park2003}, and matches the expected ratio from the natural abundances of the isotopes.

\begin{figure}[htbp]
{\includegraphics{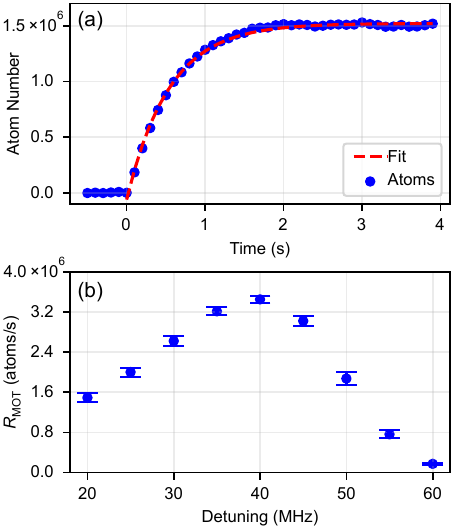}}
\caption{Loading of the Yb MOT. Plot (a): Example of a single loading run of atoms into a MOT starting at time t = 0. Oven temperature of \SI{425}{\degreeCelsius}, and detuning of \SI{45}{\mega\hertz}. Blue squares represent atom number at time t, red dashed line is the fit to loading Eq.\,\ref{eq:LR}. Plot (b): Loading rate $R_{\mathrm{MOT}}$ (atoms/s) of the MOT as a function of trapping laser detuning. Error bars represent the standard error of the mean over \num{10} runs. } 
\label{loading}
\end{figure}

Figure\,\ref{loading} shows the loading rate of the MOT as a function of detuning. These measurements were collected by analysis of MOT fluorescence images taken at \SI{50}{\milli\second} intervals as the MOT loads. The loading rate is then calculated by fitting the loading rate exponential function (Eq.\,\ref{eq:LR}) to the atom number growth curve during MOT loading:

\begin{equation}
N(t) = N_{\mathrm{eq}}\bigl(1 - e^{-t/\tau}\bigr)
\label{eq:LR}
\end{equation}

where $N(t)$ is the atom number at time $t$, $N_{\mathrm{eq}}$ is the equilibrium atom number, and $\tau$ is the loading time constant. The initial loading rate, $R_{\mathrm{MOT}} = N_{\mathrm{eq}}/\tau$, is plotted as a function of detuning in Fig.\,\ref{loading}(b). The rate peaks at $\Delta = \SI{40}{\mega\hertz}$. This optimum reflects a characteristic trade-off: while larger detunings $\Delta$ can increase the velocity capture range, this also reduces the maximum scattering force (scaling roughly as $1/\Delta^2$ for large $|\Delta|$), reducing the ability to slow atoms from the thermal beam. The observed peak in $R_{\mathrm{MOT}}$, and consequently in $N_{\mathrm{eq}}$ (Fig.\,\ref{numberDensity}), occurs where these competing factors are balanced, optimizing both the flux of captured atoms and the overall trapping efficiency.

\begin{figure}[htbp]
\includegraphics{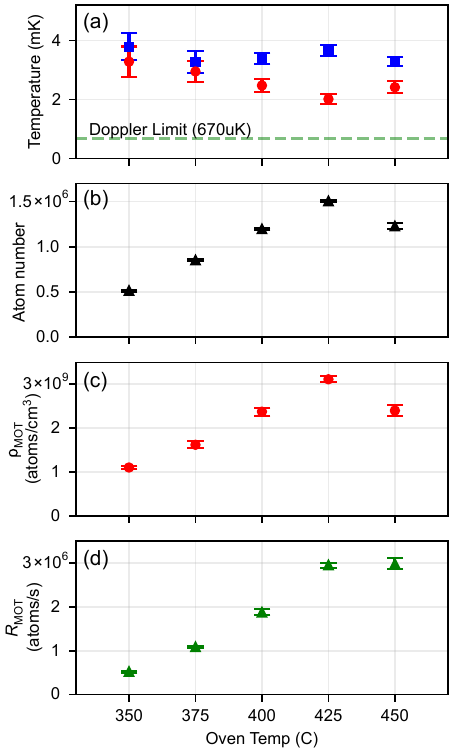}
\caption{The effect of increasing oven temperature on: (a) MOT temperature in $x$ (blue squares) and $y$ (red discs) direction (\unit{\milli\kelvin}), (b) Trapped atom number, (c) Trap density $\rho_{\mathrm{MOT}}$ (atoms/cm$^3$) , and (d) loading rate $R_{\mathrm{MOT}}$ (atoms/s). Trap detuning is fixed at \SI{45}{\mega\hertz} and magnetic field gradient  \SI{52}{G\per\centi\meter} for all measurements. For each plot, error bars represent the standard error of the mean over \num{10} runs.} 
\label{ovenTemps}
\end{figure}

Figure\,\ref{ovenTemps} illustrates the effect of oven temperature on key MOT parameters. As the oven temperature is increased from \SI{350}{\degreeCelsius} up to \SI{425}{\degreeCelsius} trapped atom number (Fig. \,\ref{ovenTemps}(b)), MOT density (Fig.\,\ref{ovenTemps}(c)), and loading rate (Fig.\,\ref{ovenTemps}(d)) all increase. This is consistent with increasing ytterbium vapor pressure, resulting in a greater atomic flux from the oven which enhances the loading of the MOT. However, beyond \SI{425}{\degreeCelsius}, these parameters begin to saturate, and a slight decrease is evident at the highest tested temperature of \SI{450}{\degreeCelsius}. This behavior indicates that at higher oven temperatures, loss mechanisms begin to counteract the benefits of the increased atomic flux. An elevated background pressure will lead to a higher rate of collisional losses for atoms trapped in the MOT, decreasing the MOT lifetime, and limiting the equilibrium trap number. The MOT is rarely used at temperatures $>$ \SI{425}{\degreeCelsius}, so this increased background pressure, which is confirmed by a rise in the ion pump pressure, can be attributed to outgassing from incompletely conditioned surfaces as the extra thermal load from heating the atomic source propagates throughout the vacuum system. In contrast, the MOT temperature (Fig.\,\ref{ovenTemps}(a)) remains relatively stable across the explored oven temperature range, suggesting that the cooling dynamics are not strongly affected by changes in atomic flux, or background pressure across this range.

\section{Discussion}

The trapped atom number results for $^{171}$Yb are comparable to earlier Yb MOTs that were loaded directly from thermal sources \cite{Rapol2004, Park2003, Xu2009}, but from a smaller sized physics package. Unsurprisingly, systems that incorporate dedicated slowing stages typically achieve significantly higher atom numbers \cite{Honda1999a, Letellier2023} resulting from enhanced MOT capture efficiency. However, the trade off in reduced atom number is a significant reduction in system size, complexity, and power requirements. The measured number of \num{4.1e6} $^{174}$Yb atoms is also consistent with expectations based on natural abundance and is comparable with other direct loading results \cite{Park2003}. The minimum temperature of \SI{680(90)}{\micro\kelvin} approaches the theoretical Doppler limit for the  \blueTransition \, transition ($\approx$\,\SI{673}{\micro\kelvin}). This is consistent with temperatures typically achieved in Yb MOTs relying primarily on Doppler cooling without dedicated sub-Doppler cooling stages, which can reach much lower temperatures \cite{Kostylev2014a, Takasu2003}. If required, lower temperatures may be possible through the use of a further detuned molasses or compression stage after initial loading of the MOT \cite{Bongs2015,Fasano2021, Zhou2013}.

The achieved atom numbers is limited by the available cooling beam power, particularly as our system operates with a saturation parameter of $s_0=0.4$. Currently there are limited options capable of providing \SI{>100}{\milli\watt} of \SI{399}{\nano\meter} light in a compact form and the cooling wavelength is on the edge of specified operating ranges for many off-the-shelf optical components, resulting in significant loss of optical power. Replacing inefficient optics could potentially double the saturation parameter to $0.8$, which is expected to increase the number of atoms within the capture range by a factor of 4 \cite{Metcalf2003}. Additionally, the atom number can be increased by the addition of a re-pump. Ref. \cite{Cho2012} was able to achieve 30\% improvement using re-pumping at \SI{649}{\nano\meter} and \SI{770}{\nano\meter}. This modest gain in performance would come at the expense of a considerable increase in SWaP and complexity to the system due to the need for additional laser sources, each of which require non-trivial spectroscopy locks. 

The optical switching mechanism for the MOT beams was significantly influenced by the operational wavelength of \SI{399}{\nano\meter} and the relatively high optical power employed (upwards of \SI{200}{\milli\watt} total). At this wavelength and power level off-the-shelf, compact AOMs with sufficient power handling are not readily available. Consequently, we chose a electro-mechanical optical shutter, which limits the switching time to \SI{700}{\micro\second}. Faster operation could have been achieved by placing the shutter at a focus in the beam, however the high power and short wavelength made this inadvisable for safety reasons.

The operational parameters of our magnetic quadrupole field, specifically its gradient, were primarily determined by the design goals of minimizing power consumption, heat dissipation, and system complexity. This resulted in an achieved gradient of \SI{52}{G\per\centi\meter} with only \SI{20}{\watt} of power dissipation. A practical constraint we observed was that changes in the drive current for the coils introduced an offset to the spatial location of the quadrupole magnetic field zero. This displacement of the MOT required realignment of the diagnostic optics for each new current setting, making systematic studies of gradient effects on MOT parameters such as atom number, temperature, and loading dynamics, difficult with the current setup. The current configuration demonstrates that a significant and stable magnetic field gradient, sufficient for reliable Yb MOT operation, can be realized with a compact, low power system. Future versions of a compact system could incorporate three-axis compensation coils to cancel out residual background fields and correct for the change in position of the zero point of the quadrupole field if fine tuning of the gradient is required for specific applications. 

The results presented here are achieved using a system built almost entirely from standard commercial components, and demonstrate an operational Yb MOT physics package within a compact SWaP footprint. Specifically, the performance of the Swagelok dual ferrule pipe fittings under UHV conditions is highly advantageous for low size and cost vacuum assembly. The wide catalog of geometries and sizes allows for relatively complex configurations while maintaining low vacuum volume. Based on the estimated flux of \SI{2.5e11}{atoms \per \second}, we expect this source to last 3 years of continuous use. Our approach thus indicates a viable pathway towards field-deployable cold Yb atom technologies. 

Substantial SWaP reduction can be expected to be reached via two different pathways. The first pathway is the implementation of custom designed elements, such as a bespoke compact vacuum system. This would allow for the placement of both the atomic source and ion pump closer to the trapping region, increasing atomic flux, and decreasing background pressures. Recent developments in pumping technology have resulted in smaller ion pumps and advances in passive pumping \cite{Kitching2018, Burrow2021}, allowing for further reduction in SWaP. The second pathway to SWaP reduction is implementation of advanced optical elements such as diffraction gratings or Fresnel reflectors. Recent work with many atomic species, specifically rubidium \cite{Nshii2013, Vangeleyn2010a, Arnold, Cotter2016}, but also cesium \cite{Yu2020, Takamizawa2024}, lithium \cite{Barker2019a, Barker2023} and strontium \cite{Bondza2022, Sitaram2020a, Barker2023}, have demonstrated grating magneto-optical traps (gMOTs) which remove the need for multiple input MOT beams, replacing them with a single input and a diffraction optic. Whilst there has been success with the group II element strontium, to date, independent trapping of Yb using a gMOT alone has not yet been achieved. Recent work has explored this area \cite{Vyalykh2024, Fasano2021, Sun2021}, with notable success with an adjacent technology: the Fresnel reflector \cite{Pick2024}. A successful implementation would result in significant reduction of system SWaP and complexity. Minimal modifications to the current system could be made to create a hot-swappable testbed for such optical technologies. Combined with the expected progress in photonic integrated circuit technology miniaturizing essential optical components such as AOMs \cite{Tian2023} and EOMs \cite{Wang2024b}, future system iterations could result in a reduction of SWaP by two or more orders of magnitude with minimal expected impact on MOT performance, thus making field-deployable ytterbium cold atom systems a reality. 

\section{Conclusion}
We have reported on a compact, low Size, Weight and Power (SWaP) Yb MOT, using a single laser system for both trapping and diagnostics, a low-complexity atomic source and a small glass spectroscopy cell. Operating on the \motTransitionBraKet \, transition, our apparatus traps \num{1.4e6} \yibbyOneSevenOne\, atoms, at a density of \num{3.1e9}{\unit{\,atoms\per\centi\meter\cubed}} and a temperature of \SI{680(90)}{\micro\kelvin}. We have shown the dependence of our MOT parameters (temperature, atom number, density, and loading rate) on cooling laser detuning and oven temperature, and outlined the challenges and possible solutions that could make deployable ytterbium cold atom technologies feasible. Looking ahead, the development of more compact designs, in particular through closer source integration and the use of single-beam technologies, will be crucial for minimizing SWaP and complexity.

\section{Acknowledgments}
This research is supported by the Commonwealth of Australia as represented by the Defense Science and Technology Group of the Department of Defence. This work was performed in part at the OptoFab node of the Australian National Fabrication Facility utilizing Commonwealth and SA State Government funding. The authors would like to thank VAF Research for their help in the fabrication of our anti-Helmholtz coils.

\bibliographystyle{apsrev4-2}
\bibliography{references}

\end{document}